# Thermal x-ray diffraction and near-field phase contrast imaging

Zheng Li[1,2,3] [(a)], Anton Classen[4,5], Tao Peng[6], Nikita Medvedev[7], Fenglin Wang[2], Henry N. Chapman[2,8,9] and Yanhua Shih[10] [(b)]

[1] *SLAC National Accelerator Laboratory, Menlo Park, California 94025, USA*
[2] *Center for Free-Electron Laser Science, DESY, Notkestrae 85, D-22607 Hamburg, Germany*
[3] *Max Planck Institute for the Structure and Dynamics of Matter, 22761 Hamburg, Germany*
[4] *Institut für Optik, Information und Photonik, Universität Erlangen-Nürnberg, 91058 Erlangen, Germany*
[5] *Erlangen Graduate School in Advanced Optical Technologies (SAOT), Universität Erlangen-Nürnberg, 91052 Erlangen, Germany*
[6] *Institute for Quantum Science and Engineering, Texas A&M University, College Station, Texas 77843, USA*
[7] *Institute of Physics and Institute of Plasma Physics, Academy of Science of Czech Republic, Na Slovance 1999/2, 18221 Prague 8, Czech Republic*
[8] *Department of Physics, University of Hamburg, Jungiusstrae 9, D-20355 Hamburg, Germany*
[9] *Hamburg Centre for Ultrafast Imaging, Luruper Chaussee 149, D-22761 Hamburg, Germany*
[10] *Department of Physics, University of Maryland, Baltimore County, Baltimore, Maryland 21250, USA*

PACS `61.05.cc` – Theories of x-ray diffraction and scattering
PACS `42.50.Ct` – Quantum description of interaction of light and matter

**Abstract** – Using higher-order coherence of thermal light sources, the resolution power of standard x-ray imaging techniques can be enhanced. In this work, we applied the higher-order measurement to far-field x-ray diffraction and near-field phase contrast imaging (PCI), in order to achieve superresolution in x-ray diffraction and obtain enhanced intensity contrast in PCI. The cost of implementing such schemes is minimal compared to the methods that achieve similar effects by using entangled x-ray photon pairs.

**Introduction.** – The Hanbury-Brown and Twiss (HBT) effect [1] was at the heart of the development of quantum optics, which fundamentally reveals the higher-order coherence of a multimode thermal source. Thermal light ghost imaging based on the HBT effect could be interpreted as a quantum mechanical two-photon interference effect [2–9]. After two decades of intense debate, the disagreement on the quantumness and classicality of thermal light ghost imaging still persists [10, 11]. However, recent rigorous investigations revealed evidence that in the low illumination (i.e. photon counting) regime, it must require a quantum model to describe the ghost image formation [12–14]. It was demonstrated that even the so-called "classical" thermal light must contain nonzero genuine quantum correlations as measured by quantum discord [15, 16]. In the large photon number regime, the ghost imaging system could transition into the classical regime, nevertheless the quantum mechanical description evidently remains valid and is quantitatively identical to the classical one, since it is the classical limit of quantum states of thermal light. Leaving aside the quantum versus classical debate, the ghost imaging actually offers enormous possibilities for x-ray imaging hitherto not fully exploited [9, 17, 18]. Here we show that using HBT type measurements, we can enhance the resolution power of conventional x-ray imaging techniques, namely x-ray far-field diffraction imaging and x-ray phase contrast imaging (PCI) in the near field. This paper is organized as follows: in the Methods section, we demonstrate how the two-photon interference effect can be utilized to x-ray diffraction imaging to double the resolution, we then investigate two-photon interference with respect to the near-field phase contrast imaging and demonstrate an enhanced image contrast in PCI. We then discuss the experimental feasibility of implementing the proposed schemes in the x-ray regime, and the methods to realize the far- and near-field imaging setups without beam splitter, which would pose technical difficulties in the x-ray regime.

[(a)]E-mail: `zheng.li@desy.de`
[(b)]E-mail: `shih@umbc.edu`

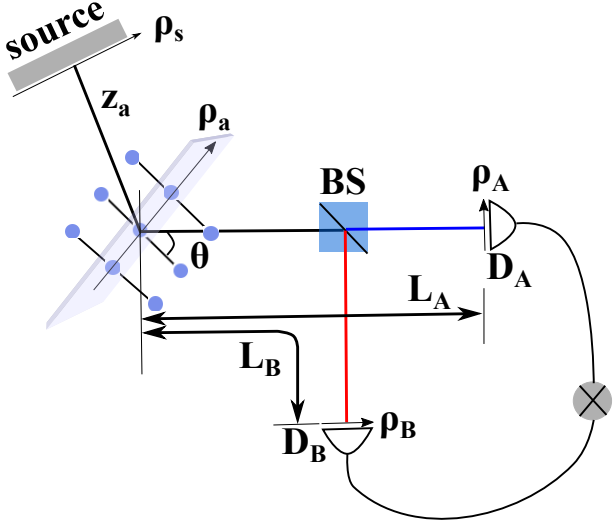

Fig. 1: Schematic description of the proposed setup for x-ray diffraction using two-photon interference. Photons from a thermal light source reflect from a crystal. The reflected light is divided by a beam splitter (BS) and detected jointly by two pixel detectors $D_A$ and $D_B$. $z_a$ is the distance between the source and lattice plane, $\theta$ is the reflection angle. $L_A$ and $L_B$ are distances from the lattice plane to the planes of pixel detectors $D_A$ and $D_B$. $\vec{\rho}_s$ is a vector on the source plane, $\vec{\rho}_a$ is a vector on the lattice plane, $\vec{\rho}_A$ and $\vec{\rho}_B$ are vectors on the detector planes.

**Methods. –**

*Enhanced x-ray diffraction.* According to the quantum theory of photo-detection [4, 5], an idealized pixel photo-detector measures the probability of observing a photo-detection event at spacetime point $(\vec{r}, t)$

$$G^{(1)}(\vec{r}, t) = \text{tr}\left\{\hat{\rho}\hat{E}^{(-)}(\vec{r}, t)\hat{E}^{(+)}(\vec{r}, t)\right\}, \quad (1)$$

where $\hat{\rho}$ is the density operator of the quantized photon field, $\hat{E}^{(-)}(\vec{r}, t)$ and $\hat{E}^{(+)}(\vec{r}, t)$ are the negative and positive frequency parts of the electric field operators. A joint-detection of two independent pixel photo-detectors $D_A$ and $D_B$ measures the probability of a coincident event of two photons at spacetime points $(\vec{r}_A, t_A)$ and $(\vec{r}_B, t_B)$

$$G^{(2)}(\vec{r}_A, t_A; \vec{r}_B, t_B) = \text{tr}\left\{\hat{\rho}\hat{E}^{(-)}(\vec{r}_A, t_A)\hat{E}^{(-)}(\vec{r}_B, t_B)\right. $$
$$\left. \times \hat{E}^{(+)}(\vec{r}_B, t_B)\hat{E}^{(+)}(\vec{r}_A, t_A)\right\}. \quad (2)$$

Let us now consider a thermal light source from which we detect the light fields at $(\vec{r}_A, t_A)$ and $(\vec{r}_B, t_B)$. This can be either a thermal light source subject to a certain propagation distance or a coherent synchrotron source. In the low photon number regime, the light field can be modeled by an effective multimode photon wavefunction $|\Psi\rangle \simeq |0\rangle + \sum_{\vec{k}} f(\vec{k})\hat{a}^\dagger_{\vec{k}}|0\rangle$, where $f(\vec{k})$ is the probability amplitude for the radiation field to be in the single-photon mode of wave vector $\vec{k}$ [2]. We assume that the light fields possess a certain coherence time $\tau_c$ over which second-order coherence is present and a time window for coincidence detection $\tau = t_2 - t_1 \ll \tau_c$. Following this assumption we can omit the time dependency from the correlation function. Note that second-order correlation function of a thermal light field measured at the same position is then given by temporal coherence of $G^{(2)}(\tau \sim 0) \sim 2$, a fact well-known as photon bunching [2]. We use the notation $\vec{r}_j = (\vec{\rho}_j, z_j)$, $(j = A, B)$ and $\vec{k} = (\vec{\kappa}, \sqrt{k^2 - \kappa^2})$ and the thermal light can be modeled as a mixed state with density operator [2]

$$\hat{\rho} \simeq |0\rangle\langle 0| + \sum_{\vec{\kappa}} |f(\vec{k})|^2 |1_{\vec{\kappa}}\rangle\langle 1_{\vec{\kappa}}|$$
$$+ \sum_{\vec{\kappa}, \vec{\kappa}'} |f(\vec{\kappa})|^2 |f(\vec{\kappa}')|^2 |1_{\vec{\kappa}} 1_{\vec{\kappa}'}\rangle\langle 1_{\vec{\kappa}} 1_{\vec{\kappa}'}|, \quad (3)$$

where $|1_{\vec{\kappa}}\rangle = \hat{a}^\dagger_{\vec{\kappa}}|0\rangle$. The transverse part of the photon field on the two detectors $D_A$ and $D_B$ can be written as

$$\hat{E}^{(+)}(\vec{\rho}_j) \simeq \sum_{\vec{\kappa}} g_j(\vec{\rho}_j, z_j; \vec{\kappa})\hat{a}_{\vec{\kappa}}, \quad (4)$$

where $g_j(\vec{\rho}_j, z_j; \vec{\kappa})$ is the Green's function for a single-mode photon. Then, the spatial part of the second-order coherence function, which is proportional to the joint photon counting rate, reads

$$G^{(2)}(\vec{\rho}_A, z_A; \vec{\rho}_B, z_B)$$
$$= \sum_{\vec{\kappa}, \vec{\kappa}'} \langle 1_{\vec{\kappa}} 1_{\vec{\kappa}'}| \hat{E}^{(-)}(\vec{\rho}_A, z_A) \hat{E}^{(-)}(\vec{\rho}_B, z_B)$$
$$\times \hat{E}^{(+)}(\vec{\rho}_B, z_A) \hat{E}^{(+)}(\vec{\rho}_A, z_B)|1_{\vec{\kappa}} 1_{\vec{\kappa}'}\rangle$$
$$= \sum_{\vec{\kappa}, \vec{\kappa}'} \left|\frac{1}{\sqrt{2}} [g_B(\vec{\rho}_B, z_B; \vec{\kappa})g_A(\vec{\rho}_A, z_A; \vec{\kappa}')\right.$$
$$\left. + g_B(\vec{\rho}_B, z_B; \vec{\kappa}')g_A(\vec{\rho}_A, z_A; \vec{\kappa})]\right|^2$$
$$\equiv G^{(1)}(\vec{\rho}_A, z_A)G^{(1)}(\vec{\rho}_B, z_B) + G^{(1)*}(\vec{\rho}_A, z_A; \vec{\rho}_B, z_B)$$
$$\times G^{(1)}(\vec{\rho}_A, z_A; \vec{\rho}_B, z_B), \quad (5)$$

where we define

$$G^{(1)}(\vec{\rho}_A, z_A; \vec{\rho}_B, z_B) = \sum_{\vec{\kappa}} g_A(\vec{\rho}_A, z_A; \vec{\kappa}) g^*_B(\vec{\rho}_B, z_B; \vec{\kappa}).$$

In Eq. 5, the term $G^{(1)}(\vec{\rho}_A, z_A)G^{(1)}(\vec{\rho}_B, z_B) \simeq \langle I(\vec{r}_A)\rangle \langle I(\vec{r}_B)\rangle$ gives a background in the joint signal intensity $\langle I(\vec{r}_A)I(\vec{r}_B)\rangle$. As shown in Fig. 2(a), the interference of a pair of independent thermal photons through two alternative yet indistinguishable paths ($|1_{\vec{\kappa}}\rangle \to D_A; |1_{\vec{\kappa}'}\rangle \to D_B$) and ($|1_{\vec{\kappa}}\rangle \to D_B; |1_{\vec{\kappa}'}\rangle \to D_A$) is manifested as joint photon number fluctuation through the term in Eq. 5 [2, 6]

$$G^{(1)*}(\vec{r}_A; \vec{r}_B)G^{(1)}(\vec{r}_A; \vec{r}_B) \simeq \langle \Delta I(\vec{\rho}_A) \Delta I(\vec{\rho}_B)\rangle. \quad (6)$$

To calculate the measurement signal quantified by Eq. 6, we propagte the thermal light field from the source to the

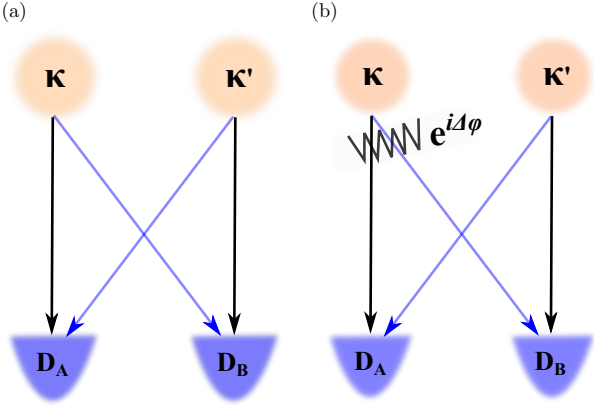

Fig. 2: (a) The two alternative yet indistinguishable paths of two single-mode photons from a thermal source, $(|1_{\vec{\kappa}}\rangle \to D_A; |1_{\vec{\kappa}'}\rangle \to D_B)$ and $(|1_{\vec{\kappa}}\rangle \to D_B; |1_{\vec{\kappa}'}\rangle \to D_A)$, the amplitudes of the two paths are added to determined the joint detection probability. (b) A phase perturbation $e^{i\Delta\varphi}$ is introduced, which models the turbulence in the experiment.

detector plane. The required Green's function reads

$$g_j(\vec{r}_i, \vec{\rho}_j, z_j; \vec{\kappa}) = \int d^2\vec{\rho}_s \int d^2\vec{\rho}_a \frac{-ik}{2\pi z_a} e^{ikz_a}$$
$$\times e^{i\frac{k}{2z_a}|\vec{\rho}_a - \vec{\rho}_s|^2} e^{i\vec{\kappa}\cdot\vec{\rho}_s} t_i(\vec{\rho}_a) \frac{-ik}{2\pi L_j} e^{ikL_j} e^{i\frac{k}{2L_j}|\vec{\rho}_j - \vec{\rho}_a|^2}, \quad (7)$$

where $j = A, B$, and $t_i(\vec{\rho}_a) = \psi\delta(\vec{\rho}_a - \vec{r}_i)$ is the Thomson scattering amplitude of a photon with the $i$th atom at position $\vec{r}_i$. Suppose the two-photon interference, which is imprinted in the photon number fluctuation, is from scattering from atoms in two adjacent lattice plane with a distance of $d$ and reflection angle $\theta$. Then the interference of the two indistinguishable two-photon quantum paths acquires a doubled optical path difference compared to conventional Laue diffraction, which translates to a maximally achievable resolution of $\frac{\lambda}{4}$ (according to Abbe's criteria for resolution and under the assumption that the measurement is taken over the full solid angle). From the Green's function $g_j(\vec{r}_i, \vec{\rho}_j, z_j; \vec{\kappa})$ for a single-mode photon that propagates to detector $D_j$ and scatters from the atom at $\vec{r}_i$, we have

$$G^{(1)*}(\vec{\rho}_A, z_A; \vec{\rho}_B, z_B)G^{(1)}(\vec{\rho}_A, z_A; \vec{\rho}_B, z_B)$$
$$= \sum_{\vec{\kappa}} \sum_{ii'}^{\text{atoms}} g_A^*(\vec{r}_i, \vec{\rho}_A, z_A; \vec{\kappa})g_B(\vec{r}_{i'}, \vec{\rho}_B, z_B; \vec{\kappa})$$
$$\times \sum_{\vec{\kappa}'} \sum_{ii'}^{\text{atoms}} g_A(\vec{r}_i, \vec{\rho}_A, z_A; \vec{\kappa}')g_B^*(\vec{r}_{i'}, \vec{\rho}_B, z_B; \vec{\kappa}'). \quad (8)$$

Taken the summation over transverse momentum as integral, and denote $\delta = d\sin\theta$, we find the following term in Eq. 8 that gives the doubled optical path difference from the two-photon interference depicted in Fig. 2(a) [19]

$$\sum_{\vec{\kappa}} g_A^*(\vec{r}_2, \vec{\rho}_A, z_A; \vec{\kappa})g_B(\vec{r}_1, \vec{\rho}_B, z_B; \vec{\kappa})$$
$$\times \sum_{\vec{\kappa}'} g_A(\vec{r}_1, \vec{\rho}_A, z_A; \vec{\kappa}')g_B^*(\vec{r}_2, \vec{\rho}_B, z_B; \vec{\kappa}') + \text{c.c.}$$
$$= \frac{32\psi^4\pi^4}{\lambda^6 L_A^2 L_B^2} \cos(4k\delta). \quad (9)$$

Conventional coherent diffraction in contrast merely yields $\cos(2k\delta)$. For x-ray diffraction, the doubled optical path difference translates to enhanced resolution, since we obtain a modified Bragg condition for the thermal light two-photon diffraction

$$4d\sin\theta = n\lambda, \quad (10)$$

for an integer $n$. It implies that structures with periods of $\frac{\lambda}{4} < d < \frac{\lambda}{2}$ are able to form Bragg peaks in the two-photon signal and can thus be resolved by inverse Fourier transformation and phase retrieval techniques. Compared to the two-photon diffraction using entangled photon pairs from x-ray parametric down conversion (XPDC), which has rather low photon flux, the present scheme using a thermal source can be experimentally favorable. In the case of powder diffraction or femtosecond serial nanocrystallography that effectively produces powder diffraction patterns, we can replace $D_A$ and $D_B$ by two pixel detectors and a sufficient large beam splitter covering the angular spread of the reflected photons, provided each pixel of the detectors are pairwise connected by a coincidence circuit. Or we can record the photon count map and correlate the image in post processing.

The thermal instability of the source, changes in refractive index in the optical paths, and shot-to-shot variation of wavefront of the x-ray beams from x-ray free electron lasers can introduce turbulence into the diffraction and imaging system. However, the two-photon interference can be free of these types of turbulence [12], hence the x-ray ghost diffraction system could be robust and does not require wavefront correction. As depicted in Fig. 2(b), the turbulence can be treated as an arbitrary phase perturbation to the Green's function in Eq. 5, and we show in the Supplementary Information, that the final diffraction pattern determined by $G^{(2)}(\vec{r}_A, \vec{r}_B)$ is invariant under phase perturbations. It can be similarly shown that the two-photon interference could be free of amplitude perturbation (see Appendix). This is a strong evidence of quantum correlation in ghost image formation, since the classical speckle-to-speckle correlation is sensitive to such turbulence [12].

*Enhanced phase contrast imaging.* As a complementary technique to the far-field x-ray diffraction, x-ray phase contrast imaging (PCI) is a widely applied near-field imaging technique which provides a direct image of the sample [20–22]. PCI is suitable for samples that only weakly absorb photons or only induce phase shift of the photon, the image of the sample is formed on a pixel detector with point-to-point projection. For the assumption that the scatterers in the sample introduce a small phase shift of $\varphi \ll \pi$ to the photon, the image is formed with irradiance that is proportional to $\nabla^2\varphi$ [22].

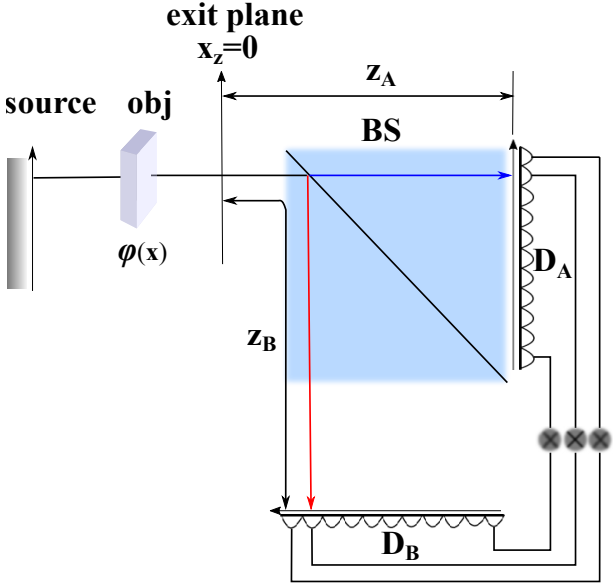

Fig. 3: The schematic configuration of phase contrast imaging using two-photon interference of thermal source. The joint coincidence measurements are performed for each pair of pixels of the two detectors $D_A$ and $D_B$. An artificial exit plane is assumed to be behind the sample, between which and the sample no further inhomogeneous phase shift $\varphi$ that has $\nabla^2\varphi \neq 0$ will be introduced. $z_i$ are the distances between the exit plane and pixel detectors.

PCI is in principle based on the near-field interference effect of an incident photon with itself that either propagates and scatters in the sample or propagates through the sample without scattering. We show here that two-photon interference in the joint detection of thermal light can be applied to PCI and enhance the intensity contrast. For the setup shown in Fig. 3, the joint intensity can be written as

$$\langle I_A(\vec{\rho}_A) I_B(\vec{\rho}_B) \rangle = G^{(2)}(\vec{\rho}_B, \vec{\rho}_A)$$
$$= G^{(1)}(\vec{\rho}_B, z_B) G^{(1)}(\vec{\rho}_A, z_A) + \left| G^{(1)}(\vec{\rho}_B, z_B, \vec{\rho}_A, z_A) \right|^2.$$

Assume the refractive index of the sample at a point $\vec{x}$ is $n_\omega(\vec{x}) = 1 - \delta_\omega(\vec{x})$ with $\delta_\omega(\vec{x}) \ll 1$. The electric field on the detector plane at spacetime $(\vec{r}_i, t)$ can be written under first Born approximation as

$$E(\vec{r}_i, \omega) = E_{\text{in}}(\vec{r}_i, \omega) - \frac{k_i^2}{2\pi} \int d^3\vec{x}' \frac{e^{ik|\vec{r}_i - \vec{x}'|}}{|\vec{r}_i - \vec{x}'|} \delta_\omega(\vec{x}') E_{\text{in}}(\vec{x}', \omega), \quad (11)$$

where $i = A, B$. Defining the coordinates on the pixel detectors as $\vec{r}_i = (\vec{\rho}_i, z_i)$, the coherence function can be expressed as

$$G^{(1)}(\vec{\rho}_B, z_B, \vec{\rho}_A, z_A)$$
$$= \frac{1}{2\pi} \iint d\omega_B d\omega_A \left\langle \hat{E}^*_{\text{in}}(\vec{\rho}_B, \omega_B) \hat{E}_{\text{in}}(\vec{\rho}_A, \omega_A) \right\rangle e^{i\Delta\omega_{AB} t}$$
$$- \frac{1}{2\pi} \iint d\omega_B d\omega_A \frac{k_A^2}{2\pi} \int d^3\vec{x}' \frac{e^{ik|\vec{r}_A - \vec{x}'|}}{|\vec{r}_A - \vec{x}'|} \delta_{\omega_A}(\vec{x}')$$
$$\times \left\langle \hat{E}^*_{\text{in}}(\vec{\rho}_B, \omega_B) \hat{E}_{\text{in}}(\vec{x}', \omega_A) \right\rangle e^{-i(\omega_B - \omega_A)t}$$
$$- \frac{1}{2\pi} \iint d\omega_B d\omega_A \frac{k_B^2}{2\pi} \int d^3\vec{x}' \frac{e^{ik|\vec{r}_B - \vec{x}'|}}{|\vec{r}_B - \vec{x}'|} \delta_{\omega_B}(\vec{x}')$$
$$\times \left\langle \hat{E}^*_{\text{in}}(\vec{x}, \omega_B) \hat{E}_{\text{in}}(\vec{r}_A, \omega_A) \right\rangle e^{-i(\omega_B - \omega_A)t}, \quad (12)$$

where $\Delta\omega_{AB} = \omega_A - \omega_B$. Under the condition $z_A = z_B$ and $\vec{\rho}_A = \vec{\rho}_B = \vec{\rho}$, we obtain the multipath coherence function as [19]

$$G^{(1)}(\vec{\rho}_B, z_B, \vec{\rho}_A, z_A) = I(t - \frac{z}{c})$$
$$\times \left( 1 - \frac{k}{4\pi z} \int d^2\vec{\rho}' l_c^2 e^{-\frac{k^2 l_c^2}{4z^2}|\vec{\rho}'|^2} \nabla^2 \varphi(\vec{\rho} - \vec{\rho}') \right).$$

The point-to-point formed image on detector plane at $\vec{\rho}$ is a convolution of the phase contrast $\nabla^2\varphi$ centered at $\vec{\rho}$ with radius $|\vec{\rho} - \vec{\rho}'|$ and a Gaussian weight. Thus the resolution of PCI is determined by requiring $e^{-\frac{k^2 l_c^2}{4z^2}p^2} \ll 1$ given $|\vec{\rho}'| = p$ as the desired resolution, i.e. the size of a single pixel. The resolution requirement leads to the condition for the coherence length

$$\frac{z\lambda}{\pi p} \ll l_c, \quad (13)$$

as well as the near-field condition $\frac{z\lambda}{p^2} \ll 1$. Provided these conditions are satisfied, we finally obtain

$$G^{(1)}(\vec{\rho}_B, z_B, \vec{\rho}_A, z_A) = I(t - \frac{z}{c}) \left( 1 - \frac{z}{k} \nabla^2 \varphi(\vec{\rho}) \right), \quad (14)$$

which has the same form as $G^{(1)}(\vec{\rho}, z)$, and does not smear out the image contrast in the baseline signal. In the conventional PCI, the image intensity is determined by [22]

$$I(\vec{\rho}, z) = G^{(1)}(\vec{\rho}, z) = I \left( 1 - \frac{z}{k} \nabla^2 \varphi(\vec{\rho}) \right). \quad (15)$$

Given a coherence length $l_c$ and incident photon wavelength, the propagation length $z$ behind the exit plane cannot be arbitrarily extended, thus in the conventional PCI, the intensity contrast $\frac{z}{k}\nabla^2\varphi$ cannot be simply enhanced. However, as we show in Eq. 14, the photon number fluctuation $\langle \Delta I_A(\vec{\rho}) \Delta I_B(\vec{\rho}) \rangle$ should not smear out the signal. Instead it has the same intensity contrast as the baseline signal $\langle I_A(\vec{\rho}) \rangle \langle I_B(\vec{\rho}) \rangle$, and thus we have the image intensity of joint photodetection as

$$\langle I_A(\vec{\rho}) I_B(\vec{\rho}) \rangle \simeq I^2 \left( 1 - \frac{2z}{k} \nabla^2\varphi(\vec{\rho}) \right), \quad (16)$$

which gives a doubled intensity contrast. It can be expected that higher order coherence measurement of thermal light [23] can further enhance the intensity contrast.

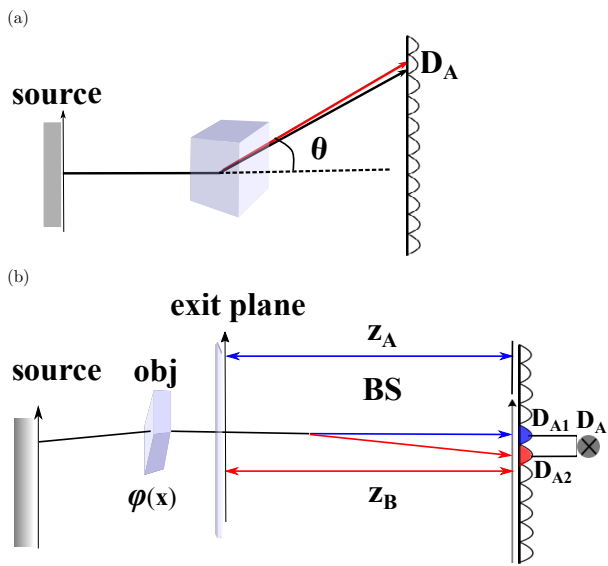

Fig. 4: The single detection configuration of (a) two-photon diffraction and (b) two-photon phase contrast imaging. In (a), the detector $D_A$ is pixellated photon counting detector, and the single pixel photon counting event of photon number $n \geq 2$ within the given time window is defined as coincidence detection. In (b), the detection of $D_{A1}$ and its adjacent pixel detector $D_{A2}$ are taken in coincidence detection as an effectively integrated single pixel $D_A$.

**Discussion.** – As shown in above derivations, the measurement of intensity-fluctuation correlations holds potential to substanially increase the performance of x-ray diffraction imaging and near-field PCI. For the practical implementation of the proposed x-ray imaging techniques based on two-photon interference, we have to consider that the use of any optical element poses a challenge toward the experimental realization. As such we discuss detection schemes here that do not require the beam splitter of Figs. 1 and 3. The beam splitter usually enables the use of a pair of single-photon counting detectors, such as avalanche photo diodes, for intensity correlation measurements. The advantages are that one can measure in CW mode, avoid dead time effects and have a high temporal resolution. However the average count rate on a single detector should be much less than unity and only a single position $\vec{\rho}_A = \vec{\rho}_B$ or $\vec{\rho}_A \neq \vec{\rho}_B$ can be measured at once. Hence the detector pair needs to be raster scanned across the detection plane.

If a pulsed scheme is considered the natural time-gating capability of ultra-short pulses removes the need for high temporal resolution in the detectors, such that a large pixelated detector can be utilized, which additionally covers the entire detection plane at once. Conducting a measurement with two different pixels $\vec{\rho}_A \neq \vec{\rho}_B$ the second-order correlation function is simply given by $G^{(2)}(\vec{\rho}_A, \vec{\rho}_B) \sim \langle \hat{a}_A^\dagger \hat{a}_B^\dagger \hat{a}_B \hat{a}_A \rangle = \langle \hat{a}_A^\dagger \hat{a}_A \hat{a}_B^\dagger \hat{a}_B \rangle = \langle \hat{n}_A \hat{n}_B \rangle$. Here $\hat{a}^\dagger$ and $\hat{a}$ denote the creation and annihilation operator of a photon in the (spatial) mode $\vec{\rho}_A$ or $\vec{\rho}_B$, respectively, and $\hat{n}$ is the corresponding number operator. Since the two pixels are independent, one can simply multiply the photon counts of different pixels and average over a series of pulses to obtain the second-order correlation function.

However, we are mostly interested in the case of $\vec{\rho}_A = \vec{\rho}_B$. In case of the PCI scheme with a continuous pattern, it is possible to correlate adjacent pixels that sample nearly the same mode created by the sample at the position of the detector (see Fig. 4b). Then the previous calculation method applies. In case of crystallography with sharp Bragg beaks (see Fig. 4a) that are much smaller than the pixel size adjacent pixel will not sample the same mode such that one would be required to use the same pixel with photon-number-resolving capability. Then $G^{(2)}(\vec{\rho}_A, \vec{\rho}_A) \sim \langle \hat{a}_A^\dagger \hat{a}_A^\dagger \hat{a}_A \hat{a}_A \rangle = \langle \hat{a}_A^\dagger (\hat{a}_A \hat{a}_A^\dagger - 1) \hat{a}_A \rangle = \langle \hat{n}_A^2 - \hat{n}_A \rangle$, which can equally be evaluated.

**Conclusion.** – In the present work, we have demonstrated that using the higher-order coherence properties of thermal light, we can achieve an enhanced resolution for conventional x-ray diffraction imaging techniques in the far field and enhanced intensity contrast in the near-field regime.

∗ ∗ ∗

The authors thank Jan Malte Slowik, Xiaolei Zhu, Matthew Ware, Robert Parrish, Andreas Kaldun, Liangliang Shi and Lida Zhang for helpful discussions. Z.L. is grateful to the Volkswagen Foundation for financial support through the Peter Paul Ewald-Fellowship. Partial financial support from the Czech Ministry of Education (Grants LG15013 and LM2015083) is acknowledged by N. Medvedev.

# epl draft

# Supplementary Information
# Thermal x-ray diffraction and near-field phase contrast imaging

Zheng Li[1,2,3], Anton Classen[4,5], Tao Peng[6], Nikita Medvedev[7], Fenglin Wang[2], Henry N. Chapman[2,8,9] and Yanhua Shih[10]

[1] *SLAC National Accelerator Laboratory, Menlo Park, California 94025, USA*
[2] *Center for Free-Electron Laser Science, DESY, Notkestrae 85, D-22607 Hamburg, Germany*
[3] *Max Planck Institute for the Structure and Dynamics of Matter, 22761 Hamburg, Germany*
[4] *Institut für Optik, Information und Photonik, Universität Erlangen-Nürnberg, 91058 Erlangen, Germany*
[5] *Erlangen Graduate School in Advanced Optical Technologies (SAOT), Universität Erlangen-Nürnberg, 91052 Erlangen, Germany*
[6] *Institute for Quantum Science and Engineering, Texas A&M University, College Station, Texas 77843, USA*
[7] *Institute of Physics and Institute of Plasma Physics, Academy of Science of Czech Republic, Na Slovance 1999/2, 18221 Prague 8, Czech Republic*
[8] *Department of Physics, University of Hamburg, Jungiusstrae 9, D-20355 Hamburg, Germany*
[9] *Hamburg Centre for Ultrafast Imaging, Luruper Chaussee 149, D-22761 Hamburg, Germany*
[10] *Department of Physics, University of Maryland, Baltimore County, Baltimore, Maryland 21250, USA*

**Abstract** – The supplementary information contains detailed derivation of main equations for thermal x-ray diffraction and near-field phase contrast imaging, more precisely, Eqs. 9 and 14 in the main text. The invariance of second order coherence function $G^{(2)}(\vec{r}_A, \vec{r}_B)$ subject to phase perturbation is also elaborated here.

**Multipath interference in diffractive imaging.** – Here we elaborate the calculation to obtain Eq. 9. We can rewrite Eq. 9 as

$$\sum_{\vec{\kappa}} g_A^*(\vec{r}_2, \vec{\rho}_A, z_A; \vec{\kappa}) g_B(\vec{r}_1, \vec{\rho}_B, z_B; \vec{\kappa})$$
$$\times \sum_{\vec{\kappa}'} g_A(\vec{r}_1, \vec{\rho}_A, z_A; \vec{\kappa}') g_B^*(\vec{r}_2, \vec{\rho}_B, z_B; \vec{\kappa}')$$
$$= G^{(1)*}(\vec{r}_2, \vec{\rho}_A.z_A; \vec{r}_1, \vec{\rho}_B, z_B)$$
$$\times G^{(1)}(\vec{r}_1, \vec{\rho}_A, z_A; \vec{r}_2, \vec{\rho}_B, z_B) \,. \tag{S.1}$$

Assume the photon scatters from atoms in two adjacent lattice planes of coordinates $\vec{0}$ and $\vec{d}$ and optical path lengths $z_a^{(1)}$ and $z_a^{(2)}$ from the source plane, the coherence functions can be calculated as

$$G^{(1)*}(\vec{r}_2, \vec{\rho}_A.z_A; \vec{r}_1, \vec{\rho}_B, z_B)$$
$$= \int d^2\vec{\kappa} \int d^2\vec{\rho}_a \int d^2\vec{\rho}_a' \frac{\psi^2}{\lambda^2 L_A L_B} \delta(\vec{\rho}_a') \delta(\vec{\rho}_a - \vec{d})$$
$$\times e^{-i\frac{k}{2L_A}|\vec{\rho}_A - \vec{\rho}_a|^2} e^{-i\frac{z_a^{(2)}}{2k}|\vec{\kappa}|^2} e^{-i\vec{\kappa}\cdot\vec{\rho}_a} e^{-ik(z_a^{(2)} + L_A)}$$
$$\times e^{i\frac{k}{2L_B}|\vec{\rho}_B - \vec{\rho}_a'|^2} e^{-i\frac{z_a^{(1)}}{2k}|\vec{\kappa}|^2} e^{i\vec{\kappa}\cdot\vec{\rho}_a'} e^{ik(z_a^{(1)} + L_B)}$$
$$= -i\frac{4\pi^2\psi^2}{\lambda^3 L_A L_B} e^{-ik(2\delta + \Delta L_{AB})} e^{-i\frac{k}{2L_A}|\vec{\rho}_A - \vec{d}|^2}$$
$$\times e^{i\frac{k}{2L_B}|\vec{\rho}_B|^2} e^{i\frac{\delta}{2k}|\vec{d}|^2} \,. \tag{S.2}$$

Similarly we can find

$$G^{(1)}(\vec{r}_1, \vec{\rho}_A.z_A; \vec{r}_2, \vec{\rho}_B, z_B) = i\frac{4\pi^2\psi^2}{\lambda^3 L_A L_B}$$
$$\times e^{-ik(2\delta - \Delta L_{AB})} e^{i\frac{k}{2L_A}|\vec{\rho}_A - \vec{d}|^2} e^{-i\frac{k}{2L_B}|\vec{\rho}_B|^2} e^{-i\frac{\delta}{2k}|\vec{d}|^2} \,.$$

And then we arrive at Eq. 9 since

$$\begin{aligned}&G^{(1)*}(\vec{r}_2,\vec{\rho}_A.z_A;\vec{r}_1,\vec{\rho}_B,z_B)\\&\times G^{(1)}(\vec{r}_1,\vec{\rho}_A,z_A;\vec{r}_2,\vec{\rho}_B,z_B)\\&=\frac{16\psi^4\pi^4}{\lambda^6 L_A^2 L_B^2}e^{-ik(4\delta)}\,.\end{aligned} \quad (S.3)$$

As illustrated in Fig. 2(b) of the main text, a phase perturbation has null effect to the second order coherence function that defines the joint photodetection intensity, because

$$\begin{aligned}&G^{(2)}(\vec{\rho}_A,z_A;\vec{\rho}_B,z_B;\Delta\varphi)\\&=\sum_{\vec{\kappa},\vec{\kappa}'}\left|\frac{1}{\sqrt{2}}\left[g_B(\vec{\rho}_B,z_B;\vec{\kappa})e^{i\Delta\varphi}g_A(\vec{\rho}_A,z_A;\vec{\kappa}')\right.\right.\\&\left.\left.+g_B(\vec{\rho}_B,z_B;\vec{\kappa}')g_A(\vec{\rho}_A,z_A;\vec{\kappa})e^{i\Delta\varphi}\right]\right|^2\\&\equiv G^{(2)}(\vec{\rho}_A,z_A;\vec{\rho}_B,z_B)\,,\end{aligned} \quad (S.4)$$

and $G^{(2)}(\vec{\rho}_A,z_A;\vec{\rho}_B,z_B)$ is thus invariant under phase perturbation [1]. It can be similarly shown that the two-photon interference could be free of amplitude perturbation. A real amplitude-wavefront distortion could be introduced into the model of ghost imaging system by replacing the phase perturbation $e^{i\Delta\varphi}$ in Eq. S.4 with an amplitude perturbation $\Delta E$. In this case, the resulting second order coherence function $G^{(2)}(\vec{\rho}_A,z_A;\vec{\rho}_B,z_B)$ acquires merely a prefactor and has no effect on the visibility and spatial frequencies of the image.

The turbulence-free property of the ghost imaging system can be considered as an evidence for quantum correlation of two-photon interference. In the classical simulation of ghost imaging, the speckle-to-speckle correlation was used, and the speckles of the thermal source is imaged onto the object plane and the image plane [2–4]. The two sets of identical intensity speckles $\langle I_A\rangle\langle I_B\rangle$ are measured and correlated to generate image of the object in the classical simulation. A genuine quantum ghost image is determined from the intensity fluctuation correlation $\langle\Delta I_A\Delta I_B\rangle$ by photon counting, which has a relation with the classical simulation as

$$\langle I_A I_B\rangle = \langle I_A\rangle\langle I_B\rangle + \langle\Delta I_A\Delta I_B\rangle\,, \quad (S.5)$$

and reflects the two-photon interference effect, i.e. a pair of photons interfering with the pair itself. It was demonstrated experimentally that the ghost image observed from $\langle\Delta I_A\Delta I_B\rangle$ is turbulence-free. However, the image from classical simulation of $\langle I_A\rangle\langle I_B\rangle$ is turbulence-sensitive [1], because the corresponding light intensity $|\sum_{\vec{\kappa}}g(\vec{\rho},z;\vec{\kappa})e^{i\Delta\varphi}|^2$ is not invariant under perturbation.

**Multipath interference in phase contrast imaging.** – The major task of this section is to elaborate the calculation to obtain Eq. 14 of the main text. In Eq. S.7, the coherence function describes self-interference of the incident photon that either scatters from the sample or directly transverses through the sample. The electric field of the photon satisfies the Helmholtz equation [5]

$$\left(\nabla^2+k^2\right)E(\vec{r}_i,\omega)=k^2\left(1-n_\omega^2(\vec{r}_i)\right)E(\vec{r}_i,\omega) \quad (S.6)$$

From Eq. S.6, the electric field on the detector plane at spacetime $(\vec{r}_i,t)$ can be written under first Born approximation as

$$\begin{aligned}E(\vec{r}_i,\omega)&=E_{\rm in}(\vec{r}_i,\omega)\\&-\frac{k_i^2}{2\pi}\int d^3\vec{x}'\frac{e^{ik|\vec{r}_i-\vec{x}'|}}{|\vec{r}_i-\vec{x}'|}\delta_\omega(\vec{x}')E_{\rm in}(\vec{x}',\omega)\,,\end{aligned}$$

where $i=A,B$. Defining the coordinates on the pixel detectors $\vec{r}_i=(\vec{\rho}_i,z_i)$, the coherence function can be expressed as Eq. 12 in the main text

$$\begin{aligned}&G^{(1)}(\vec{\rho}_B,z_B,\vec{\rho}_A,z_A)\\&=\frac{1}{2\pi}\iint d\omega_B d\omega_A\left\langle\hat{E}_{\rm in}^*(\vec{\rho}_B,\omega_B)\hat{E}_{\rm in}(\vec{\rho}_A,\omega_A)\right\rangle e^{i\Delta\omega_{AB}t}\\&-\frac{1}{2\pi}\iint d\omega_B d\omega_A\frac{k_A^2}{2\pi}\int d^3\vec{x}'\frac{e^{ik|\vec{r}_A-\vec{x}'|}}{|\vec{r}_A-\vec{x}'|}\delta_{\omega_A}(\vec{x}')\\&\times\left\langle\hat{E}_{\rm in}^*(\vec{\rho}_B,\omega_B)\hat{E}_{\rm in}(\vec{x}',\omega_A)\right\rangle e^{-i(\omega_B-\omega_A)t}\\&-\frac{1}{2\pi}\iint d\omega_B d\omega_A\frac{k_B^2}{2\pi}\int d^3\vec{x}'\frac{e^{ik|\vec{r}_B-\vec{x}'|}}{|\vec{r}_B-\vec{x}'|}\delta_{\omega_B}(\vec{x}')\\&\times\left\langle\hat{E}_{\rm in}^*(\vec{x},\omega_B)\hat{E}_{\rm in}(\vec{r}_A,\omega_A)\right\rangle e^{-i(\omega_B-\omega_A)t}\\&\equiv I_{(0,0)}-I_{(0,1)}-I_{(1,0)}\,,\end{aligned} \quad (S.7)$$

where $\Delta\omega_{AB}=\omega_A-\omega_B$. Defining the cross spectral density [6]

$$W(\vec{r}_1,\vec{r}_2,\omega_1,\omega_2)=\left\langle\hat{E}_{\rm in}^*(\vec{r}_1,\omega_1)\hat{E}_{\rm in}(\vec{r}_2,\omega_2)\right\rangle\,, \quad (S.8)$$

we can rewrite the $I_{(0,1)}$ term in Eq. S.7 as

$$\begin{aligned}I_{(0,1)}(\vec{r}_B,\vec{r}_A)&=\frac{1}{4\pi^2}\iint\omega_B\omega_A\int d^3\vec{x}'k_A^2\frac{e^{ik_A|\vec{r}_A-\vec{x}'|}}{|\vec{r}_A-\vec{x}'|}\\&\times\delta_{\omega_A}(\vec{x}')W(\vec{r}_B,\vec{x}',\omega_B,\omega_A)e^{-i(\omega_B-\omega_A)t}\,.\end{aligned} \quad (S.9)$$

Denote $\omega=\frac{\omega_A+\omega_B}{2}$, $\overline{\omega}=\omega_A-\omega_B$, using the Huygens principle and the Rayleigh-Sommerfeld diffraction integral of the first kind for propagating the cross spectral density from the exit plane to the detector plane [5], we have

$$\begin{aligned}&W(\vec{r}_B,\vec{x}',\omega,\overline{\omega})\\&=-\frac{1}{2\pi}\int_{x_z=0}d^2\vec{x}W(\vec{x},\vec{x}',\omega,\overline{\omega})\frac{\partial}{\partial z}\left(\frac{e^{-ik_B|\vec{r}_B-\vec{x}|}}{|\vec{r}_B-\vec{x}|}\right)\,,\end{aligned}$$

where $\vec{x}=(\vec{x}_\perp,x_z=0)$ defines the coordinate on the exit plane. The sample is assumed to be in front of the exit plane with $x_z<0$, such that after passing the exit plane, no inhomogeneous phase shift will be introduced. Introducing the coherence length $l_c$ of the incident photon beam, the cross spectral density $W(\vec{x},\vec{x}',\omega,\overline{\omega})$ has the form [6,7]

$$W(\vec{x},\vec{x}',\omega,\overline{\omega})=2\pi\gamma(\vec{x},\vec{x}')I\delta(\overline{\omega})\delta(\omega-\omega_{\rm in})\,, \quad (S.10)$$

where $\gamma(\vec{x}, \vec{x}') = e^{-|\vec{x}_\perp - \vec{x}'_\perp|^2/l_c^2}$ describes the transverse coherence between two points on the exit plane, a large coherence length $l_c$ is crucial to the resolution of PCI. For thermal sources, we have $l_c = \frac{\lambda D}{a}$, where $\lambda$ is the photon wavelength, $D$ is the distance between the source and sample, and $a$ is the source radius, it is always possible to obtain sufficient coherence length by taking a large propagation distance $D$. With Eqs. S.10 and S.10, we can further obtain

$$I_{(0,1)} = -\frac{1}{8\pi^3} \iint d\omega d\overline{\omega} \int d^3\vec{x}' \alpha^2(\omega + \overline{\omega}/2)^2$$
$$\times \frac{e^{i\alpha(\omega+\overline{\omega}/2)|\vec{r}_A - \vec{x}'|}}{|\vec{r}_A - \vec{x}'|} \delta_{\omega+\overline{\omega}/2}(\vec{x}') \int_{x_z=0} d^2\vec{x}$$
$$W(\vec{x},\vec{x}',\omega,\overline{\omega}) \frac{\partial}{\partial z}\left(\frac{e^{-i\alpha(\omega-\overline{\omega}/2)|\vec{r}_B - \vec{x}|}}{|\vec{r}_B - \vec{x}|}\right), \quad (S.11)$$

where $\alpha = 1/c$ is the fine structure constant. Denote the coordinate between the exit plane and the detector as $\vec{r} = (\vec{\rho}, z)$, we use the relation $\frac{e^{-ik|\vec{r}-\vec{x}|}}{|\vec{r}-\vec{x}|} \simeq \frac{e^{-ik(z+|\vec{\rho}-\vec{x}_\perp|^2/2z)}}{z}$ and obtain

$$I_{(0,1)} = -\frac{1}{8\pi^3} \iint d\omega d\overline{\omega} \int d^3\vec{x}' \alpha^2(\omega+\overline{\omega}/2)^2$$
$$\times \frac{e^{i\alpha(\omega+\overline{\omega}/2)(z_A + |\vec{\rho}_A - \vec{x}'_\perp|^2/2z_A)}}{z_A} \delta_{\omega+\overline{\omega}/2}(\vec{x}') \int_{x_z=0} d^2\vec{x}$$
$$2\pi I \gamma(\vec{x}_\perp, \vec{x}'_\perp) \delta(\omega - \omega_{\rm in})\delta(\overline{\omega})[-i\alpha(\omega-\overline{\omega}/2)]$$
$$\times \frac{e^{-i\alpha(\omega-\overline{\omega}/2)(z_B + |\vec{\rho}_B - \vec{x}'_\perp|^2/2z_B)}}{z_B} e^{i\overline{\omega}t}$$
$$= iI\frac{k^2}{4\pi^2 z_A z_B} \int d^2\vec{x}'_\perp \int_{x_z=0} d^2\vec{x}_\perp \gamma(\vec{x}_\perp, \vec{x}'_\perp)$$
$$\times \left[k \int dx'_z \delta_\omega(\vec{x}')\right] e^{ik\left(\frac{|\vec{\rho}_A - \vec{x}'_\perp|^2}{2z_A} - \frac{|\vec{\rho}_B - \vec{x}_\perp|^2}{2z_B}\right)} e^{ik(z_A - z_B)},$$

where we denote $\omega \equiv \omega_{\rm in}$. The accumulated phase shift for a photon that travels through the sample and arrives at the exit plane is defined as

$$\varphi(\vec{x}'_\perp) = -k \int dx'_z \delta_\omega(\vec{x}'), \quad (S.12)$$

with which we write $I_{(0,1)}$ as

$$I_{(0,1)} = -iI\frac{k^2}{4\pi^2 z^2} \int d^2\vec{x}'_\perp \int d^2\vec{x}_\perp \gamma(\vec{x}_\perp, \vec{x}'_\perp)\varphi(\vec{x}'_\perp)$$
$$\times e^{i\frac{k}{2}\left(\frac{|\vec{\rho}_A - \vec{x}'_\perp|^2}{z_A} - \frac{|\vec{\rho}_B - \vec{x}_\perp|^2}{z_B}\right)} e^{ik(z_A - z_B)}, \quad (S.13)$$

similarly we found the form of $I_{(1,0)}$ in Eq. S.7. As shown in Fig. 3 of the main text, the pixels of the two detectors are pairwise connected through coincidence circuit, we can thus assume $\vec{\rho}_A = \vec{\rho}_B \equiv \vec{\rho}$. We denote

$$\frac{z_B - z_A}{z_A z_B} = \frac{-\eta}{z}$$
$$\frac{z_B + z_A}{z_A z_B} = \frac{2\zeta}{z}$$
$$z_{AB} = z_A - z_B, \quad (S.14)$$

where for the case $z_A = z_B = z$, we have $\eta = 0, \zeta = 1, z_{AB} = 0$. The interference kernel of the coherence function is thus

$$I_{(0,1)} + I_{(1,0)}$$
$$= -iI\frac{k^2}{4\pi^2 z^2} \int d^2\vec{a} \int d^2\vec{b} \left[e^{-\left(\frac{1}{l_c^2} + i\frac{k\eta}{8z}\right)b^2}\right]$$
$$\times e^{-i\frac{\eta k}{4z}a^2 + i\frac{\eta k}{2z}\vec{\rho}\cdot\vec{a}} \varphi(\vec{a} - \vec{b}/2) \left[e^{-i\frac{\zeta k}{z}(\vec{a}-\vec{\rho})\cdot\vec{b}} e^{-ikz_{AB}}\right.$$
$$\left. - e^{i\frac{\zeta k}{z}(\vec{a}-\vec{\rho})\cdot\vec{b}} e^{ikz_{AB}}\right], \quad (S.15)$$

where $\vec{a} = \frac{1}{2}(\vec{x}_\perp + \vec{x}'_\perp)$ and $\vec{b} = \vec{x}_\perp - \vec{x}'_\perp$. Further taken transformation $\vec{a} \equiv \vec{a} - \vec{b}/2$, we have

$$I_{(0,1)} + I_{(1,0)}$$
$$= -iI\frac{k^2}{4\pi^2 z^2} \int d^2\vec{b} \int d^2\vec{a} e^{-i\left(\frac{1}{l_c^2} + i\frac{k\eta}{8z}\right)b^2} \varphi(\vec{a})$$
$$\times \left\{ e^{-i\frac{\eta k}{4z}[(\vec{a}+\vec{b}/2+\vec{\rho})^2 - \rho^2]} e^{i\frac{\zeta k}{z}(\vec{a}\cdot\vec{b} - \vec{\rho}\cdot\vec{b}) - ikz_{AB}} e^{i\frac{\zeta k}{2z}b^2} \right.$$
$$\left. - e^{-i\frac{\eta k}{4z}[(\vec{a}-\vec{b}/2+\vec{\rho})^2 - \rho^2]} e^{i\frac{\zeta k}{z}(\vec{a}\cdot\vec{b} - \vec{\rho}\cdot\vec{b}) + ikz_{AB}} e^{-i\frac{\zeta k}{2z}b^2} \right\}.$$

Taken the simplest case $z_A = z_B$, we have

$$I_{(0,1)} + I_{(1,0)}$$
$$= I\frac{k^2}{2\pi^2 z^2} \int d^2\vec{b} \int d^2\vec{a} e^{-\frac{b^2}{l_c^2}} \varphi(\vec{a}) e^{-i\frac{k}{z}\vec{\rho}\cdot\vec{b}}$$
$$\times e^{i\frac{k}{z}\vec{b}\cdot\vec{a}} \sin\left(\frac{k}{2z}b^2\right)$$
$$\simeq I\frac{k^2}{\pi z^2} \int d^2\vec{b} e^{-\frac{b^2}{l_c^2}} [\mathcal{F}\varphi]\left(\frac{k}{z}\vec{b}\right) \frac{k}{2z} b^2$$
$$\times e^{-i\frac{k}{z}\vec{\rho}\cdot\vec{b}}. \quad (S.16)$$

Defining transformation $\vec{b} = \frac{k}{z}\vec{b}$,

$$I_{(0,1)} + I_{(1,0)}$$
$$= I\frac{z}{2\pi k} \int d^2\vec{b} e^{-\frac{z^2 b^2}{k^2 l_c^2}} [\mathcal{F}\nabla^2\varphi](\vec{b}) e^{-i\vec{\rho}\cdot\vec{b}}$$
$$= I\frac{z}{4\pi^2 k} \left\{ \mathcal{F}^{-1}\left(e^{-\frac{z^2 b^2}{k^2 l_c^2}}\right) \otimes \mathcal{F}^{-1}\left[\mathcal{F}\nabla^2\varphi\right]\right\}(\vec{\rho})$$
$$= I\frac{z}{2\pi k} \int d^2\vec{\rho}' \frac{k^2 l_c^2}{2z^2} e^{-\frac{k^2 l_c^2}{4z^2}|\vec{\rho}'|^2} \nabla^2\varphi(\vec{\rho} - \vec{\rho}'). \quad (S.17)$$

We define $\vec{y}' = \frac{kl_c}{2z}\vec{\rho}'$ and obtain

$$I_{(0,1)} + I_{(1,0)}$$
$$= I\frac{z}{\pi k} \int d^2\vec{y}' e^{-y'^2} \nabla^2\varphi\left(\vec{\rho} - \frac{2z}{kl_c}\vec{y}'\right). \quad (S.18)$$

For sufficiently large $l_c$, which guarantees rapid decaying of the convolution factor that satisfies $e^{-y'^2} \ll 1$ with given pixel size (resolution) $|\vec{\rho}'| = p$, Eq. S.18 reduces to

$$I_{(0,1)} + I_{(1,0)} = I\frac{z}{k} \nabla^2\varphi(\vec{\rho}), \quad (S.19)$$

And Eq. 14 of the main text follows as a result.